\documentclass[fleqn,10pt]{wlscirep}
\usepackage[utf8]{inputenc}
\usepackage[T1]{fontenc}
\usepackage{multirow}

\title{Event topology and global observables in heavy-ion collisions at the Large Hadron Collider}

\author[1]{Suraj Prasad}
\author[1]{Neelkamal Mallick}
\author[1]{Debadatta Behera}
\author[1,2,*]{Raghunath Sahoo}
\author[3]{Sushanta Tripathy}

\affil[1]{Department of Physics, Indian Institute of Technology Indore, Simrol, Indore 453552, India}
\affil[2]{CERN, CH 1211, Geneva 23, Switzerland}
\affil[3]{INFN - sezione di Bologna, via Irnerio 46, 40126 Bologna BO, Italy}

\affil[*]{Corresponding Author Email: Raghunath.Sahoo@cern.ch}

\begin{abstract}
Particle production and event topology are very strongly correlated in high-energy hadronic and nuclear collisions. Event topology is 
decided by the underlying particle production dynamics and medium effects. Transverse spherocity is an event shape observable, 
which has been used in pp  and heavy-ion collisions to separate the events  based on their geometrical shapes. It has the unique 
capability to distinguish between jetty and isotropic events. In this work, we have implemented transverse spherocity in Pb-Pb collisions at 
$\sqrt{s_{\rm NN}}$ = 5.02 TeV using A Multi-Phase Transport Model (AMPT). While awaiting for experimental explorations, we perform 
a feasibility study of dependence of transverse spherocity on some of the global observables in heavy-ion collisions at the Large Hadron Collider energies. These global observables include the Bjorken energy density ($\epsilon_{\rm Bj}$), squared speed of sound ($c_{\rm s}^2$) in the medium and the kinetic freeze-out properties for different collision centralities. The present study reveals about the usefulness of 
event topology dependent measurements in heavy-ion collisions.
\end{abstract}
\begin{document}

\flushbottom
\maketitle
%
%
\thispagestyle{empty}


\section{Introduction}
\label{intro}

Heavy-ion collisions at the ultra-relativistic energies aim to produce a deconfined state of quarks and gluons, the primordial matter
believed to have formed at the infancy of the Universe. The matter created in such collisions at the Relativistic Heavy-Ion Collider
 (RHIC) at the Brookhaven National Laboratory, USA and at the Large Hadron Collider (LHC) at European Center for Nuclear Research (CERN), Switzerland
 gives an opportunity to study its properties at the extreme conditions of temperature and energy densities. Global properties
 of the created matter such as total charged particle multiplicity in the final state, initial energy density and temperature of the system
 play a pivotal role to understand the form of the created matter, while addressing many fundamental questions in basic science.
 The expansion of the created fireball because of huge concentration of initial energy density and high temperature is probed
 through the equation of state and hence the speed of sound in the medium. The final state particle abundances are governed
 by the initial state of the matter -- if the collision creates a partonic deconfined colored phase of matter or a confined, color neutral
  hadronic state. The event topology is governed by the underlying particle production mechanism. For example, a back-to-back
  momentum conserving shower of particles, called jets is an event topology, whose underlying mechanism is governed by 
  hard perturbative Quantum Chromodynamics (pQCD) high transverse momentum processes, whereas an isotropic geometry in the
  event topology is mostly rich in soft QCD non-perturbative interactions.

\begin{figure}[ht]
\begin{center}
\includegraphics[scale=0.4]{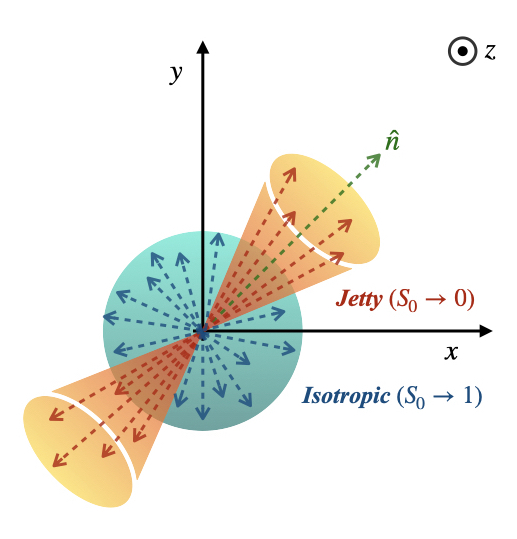}
\caption[]{(Color Online) Schematic picture showing jetty and isotropic events in the transverse plane, assuming the $z$-axis is the beam axis or the longitudinal axis.}
\label{sp_cart}
\end{center}
\end{figure}

In elementary and hadronic collisions at GeV and TeV energies, although event topology dependent studies have got some 
level of importance, because of a dense medium formation in heavy-ion collisions, these techniques like sphericity, transverse spherocity,
$R_T$ etc. are not applied to heavy-ion collisions. Event topology dependent characterisation of the systems produced in heavy-ion collisions through the global properties are not a well explored area. This is a first attempt to explore the sensitiveness of the global observables in heavy-ion collisions, to the event topology and hence the underlying particle production dynamics. In our recent study~\cite{Mallick:2020ium}, we found that the anisotropic flow strongly depends on transverse spherocity in heavy-ion collision systems. Thus, it would be interesting to see how the global properties, often studied in heavy-ion collisions, vary as a function of transverse spherocity. It is worth to note that global properties like Bjorken energy density and speed of sound give insights to the initial state of the produced system while the kinetic freezeout parameters provide insights to the evolution of produced particles in the medium. The use of transverse spherocity along with collision centrality also provides an opportunity to study such observables in a multi-differential way. In the present work, we use transverse spherocity as the event shape observable and study the global properties such as Bjorken energy density, speed of sound and kinetic freeze-out parameters for different centrality classes in Pb-Pb collisions at $\sqrt{s_{\rm NN}}$ = 5.02 TeV using A Multi-Phase Transport Model (AMPT).

The paper is organised as follows. We begin with a brief introduction about the usefulness of event topology studies in heavy-ion collisions. In section~\ref{section2}, the event generation methodology in AMPT and the definition of transverse spherocity is given. We report and discuss the results in section~\ref{section3}. The results are summarised in section~\ref{section4}.

\section{Event Generation and Analysis Methodology}
\label{section2}

In this section, we begin with a brief introduction on AMPT model. Then, we proceed to define the transverse spherocity as an event shape analysis tool. 

\subsection{A Multi-Phase Transport (AMPT) Model}
\label{formalism}
A Multi-Phase Transport Model contains four components namely~\cite{AMPT2,ampthijing,amptzpc,amptreco,Greco:2003mm,amptart1, amptart2,ampthadron1,ampthadron2,ampthadron3}, 
\begin{itemize}
\item Initialisation of collisions using HIJING model: the cross-section of the produced mini-jets in pp collisions is calculated and then converted to heavy-ion collisions via inbuilt Glauber model 
\item Parton transport after initialisation:  transportation of produced partons is performed via Zhang’s parton cascade model
\item Hadronisation mechanism: in string melting version, the transported partons are hadronised using spatial coalescence mechanism; in the default AMPT version, fragmentation mechanism using Lund fragmentation parameters are used for hadronising the transported partons
\item Hadron transport: the hadrons undergo evolution in relativistic transport mechanism via meson-baryon, meson-meson and baryon-baryon interactions
\end{itemize}

As, the particle flow and spectra at the mid-$p_{\rm T}$ regions are well explained by quark coalescence mechanism for hadronisation~\cite{ampthadron1,ampthadron2,ampthadron3}, we have used string melting mode for all of our calculations. We have used the AMPT version 2.26t7 (released: 28/10/2016) in our current work. The AMPT settings in the current work, are the same as reported in Ref.~\cite{Tripathy:2018bib,Mallick:2020ium}. For the input of impact parameter values for different centralities in Pb-Pb collisions, we have used Ref.~\cite{Loizides:2017ack}. One should note here that, high centrality collisions correspond to low impact parameter values and higher final state charged-particle multiplicity ($\langle dN_{\rm ch}/d\eta \rangle$). Although the concept of centrality is widely used in heavy-ion collisions, in view of a final state multiplicity scaling across collisions species, that is observed at the LHC energies, we may use centrality and $\langle dN_{\rm ch}/d\eta \rangle$ variably in this work.

\subsection{Transverse Spherocity}

\begin{figure}[ht]
\begin{center}
\includegraphics[scale=0.4]{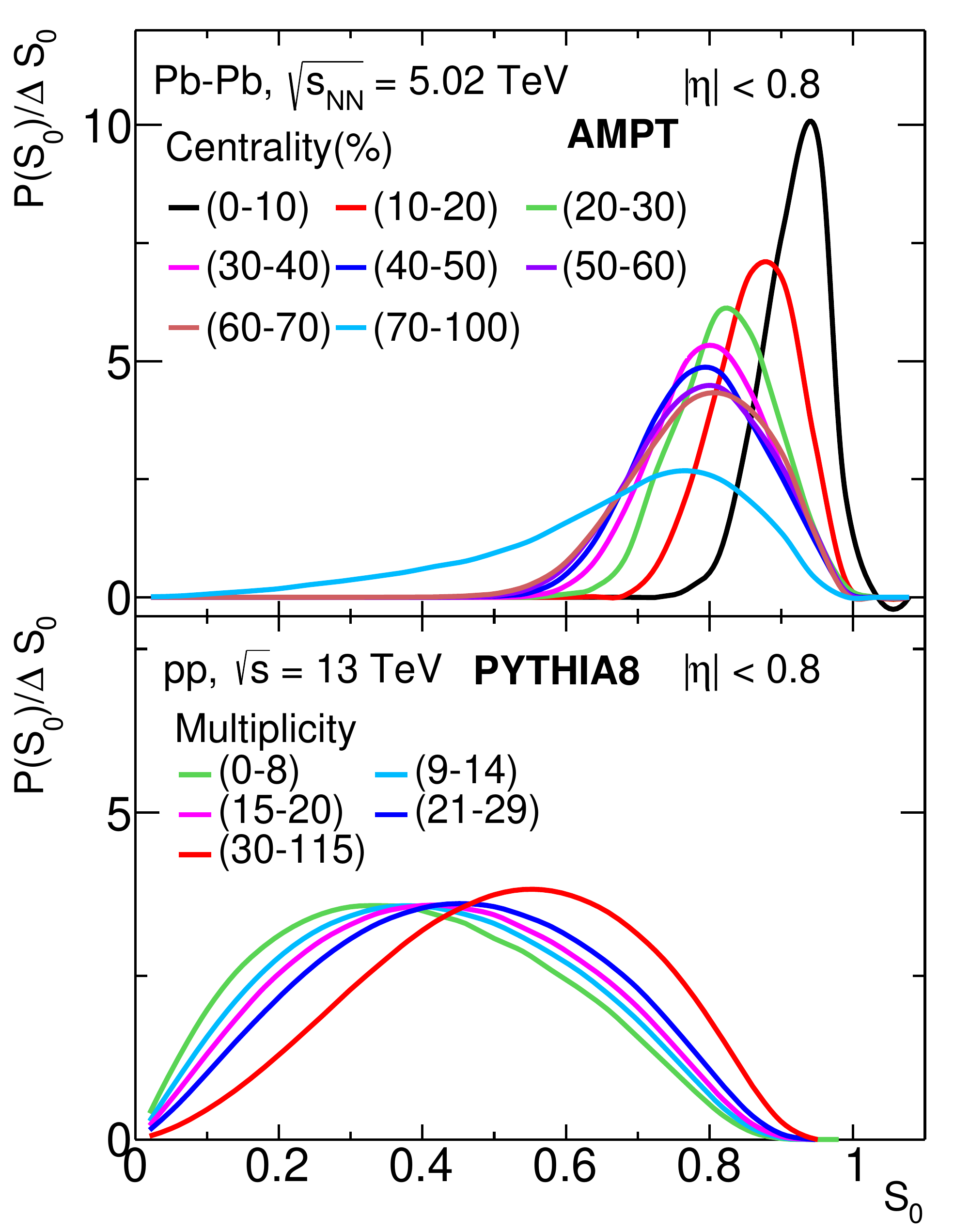}
\caption[]{(Color Online) Top (bottom) panel: transverse spherocity distribution for different centrality (multiplicity) classes in Pb-Pb (pp) collisions at $\sqrt{s_{\rm NN}}=5.02 ~\rm{TeV}$ ($\sqrt{s}=13 ~\rm{TeV}$) using AMPT (PYTHIA8) model.}
\label{SpheroDist}
\end{center}
\end{figure}

Transverse spherocity is defined for a unit vector $\hat{n} (n_{T},0)$ that minimizes the ratio~\cite{Cuautle:2014yda, Cuautle:2015kra}:
\begin{eqnarray}
S_{0} = \frac{\pi^{2}}{4} \bigg(\frac{\Sigma_{i}~|\vec p_{T_{i}}\times\hat{n}|}{\Sigma_{i}~p_{T_{i}}}\bigg)^{2}.
\label{eq7}
\end{eqnarray}

$\hat{n}$ is an arbitrary unit vector in the transverse plane. To find such a unit vector, one has to perform iteration through all possible values of $\hat{n}$ with azimuthal angle $0$ to $2\pi$ and select the $\hat{n}$ such that the term inside the bracket given in Eq. \ref{eq7} becomes minimum for a given event. Here, the index $i$ runs over all the final state hadrons in an event. By construction, transverse spherocity is infrared and collinear safe~\cite{Salam:2009jx} and the extreme limits are related to specific configurations of events in transverse plane. Transverse spherocity becoming 0 are the events with pencil-like (back-to-back) structure and called as jetty events while 1 would mean the events are isotropic. A schematic picture illustrating the event topology is shown in Fig.~\ref{sp_cart}. For the sake of simplicity, here onwards, the transverse spherocity is referred as spherocity. The spherocity distributions are selected in the pseudorapidity range of $|\eta|<0.8$ with a minimum constraint of 5 charged particles with $p_{\rm{T}}>$~0.15~GeV/$c$ to recreate the similar conditions as in ALICE experiment at the LHC. The jetty events are those events having spherocity values in the lowest 20 percent and the isotropic events are those occupying the highest 20 percent in the spherocity distribution of all the events. The spherocity cuts for each centrality are mentioned in Table \ref{tab:1}.

\begin{table}[ht!]
\begin{center}
\begin{tabular}{ |p{2cm}|p{2cm}|p{2cm}|}
 \hline
 Centrality (\%) & Low-$S_{0}$ & High-$S_{0}$\\
\hline
0-10         & 0 -- 0.880         & 0.953 -- 1 \\
10-20         & 0 -- 0.813        &    0.914 -- 1 \\
20-30 & 0 -- 0.760  & 0.882 -- 1 \\
30-40 & 0 -- 0.735  & 0.869 -- 1 \\
40-50 & 0 -- 0.716  & 0.865 -- 1 \\
50-60 & 0 -- 0.710  & 0.870 -- 1 \\
60-70 & 0 -- 0.707  & 0.873 -- 1 \\
70-100 & 0 -- 0.535  & 0.822 -- 1 \\
 \hline
 \end{tabular}
 \end{center}
 \caption{Low 20 \% and high 20\% cuts on spherocity distribution in Pb-Pb collisions at $\sqrt{s_{\rm{NN}}} = 5.02$~TeV for different centrality classes. }
\label{tab:1}
\end{table}

Top panel of Fig.~\ref{SpheroDist} shows the spherocity distributions for different centrality classes in Pb-Pb collisions, $\sqrt{s_{\rm NN}}=5.02 ~\rm{TeV}$ at mid-rapidity ($|\eta| < 0.8$) using AMPT model and similarly, the bottom panel is the spherocity distribution for different charged particle multiplicity classes in pp collisions, $\sqrt{s}=13 ~\rm{TeV}$ at mid-rapidity ($|\eta| < 0.8$) using PYTHIA8~\cite{Sjostrand:2014zea}. The details of event generation methodology using PYTHIA8 can be found in Ref. \cite{Khuntia:2018qox}. For pp collisions, the charged particle multiplicities are chosen in the acceptance of V0 detector in ALICE at the LHC with pseudorapidity coverage of V0A ($2.8< \eta <5.1$) and V0C ($ -3.7< \eta < -1.7$). One should note here that the particle production mechanisms in AMPT and PYTHIA8 models are completely different but we have chosen the tunes of the models where the models describe many of the experimental observables. At a first glance, the spherocity distributions in Pb-Pb collisions look shifted more towards the isotropic limit when compared to the pp collisions, where the distributions are shifted towards the jetty limit. This behavior is understood based on the fact that the system size in Pb-Pb collisions are significantly higher when compared with pp collisions and the medium effect in terms of rescattering helps taking the system towards isotropisation. In comparison to pp collisions, where we observe a distribution of spherocity or in other words, there is an equal production probability of both jetty and isotropic events, medium effects
in heavy-ion collisions in principle destroy the jettiness event topology, which is seen from Fig.~\ref{SpheroDist}. A skewed distribution of spherocity towards isotropic limit is an indication of the formation of a QCD medium in heavy-ion collisions. When studied as a function of centrality (multiplicity) classes for Pb-Pb (pp) collisions, it is observed that the spherocity distributions are shifted towards the isotropic limit for central (high-multiplicity) collisions compared to peripheral (low-multiplicity) collisions.

Now, we proceed to discuss the global properties and their dependence on spherocity classes in the next section. For the sake of simplicity, here onwards we refer $\pi^{+}+\pi^{-}$, K$^{+}$+K$^{-}$, and p+$\bar{\rm p}$ as pions, kaons, and protons, respectively.

\section{Results and Discussions}
\label{section3}

\subsection{Bjorken Energy density ($\epsilon_{\rm Bj})$}

\begin{figure}[ht!]
\begin{center}
\includegraphics[scale=0.44]{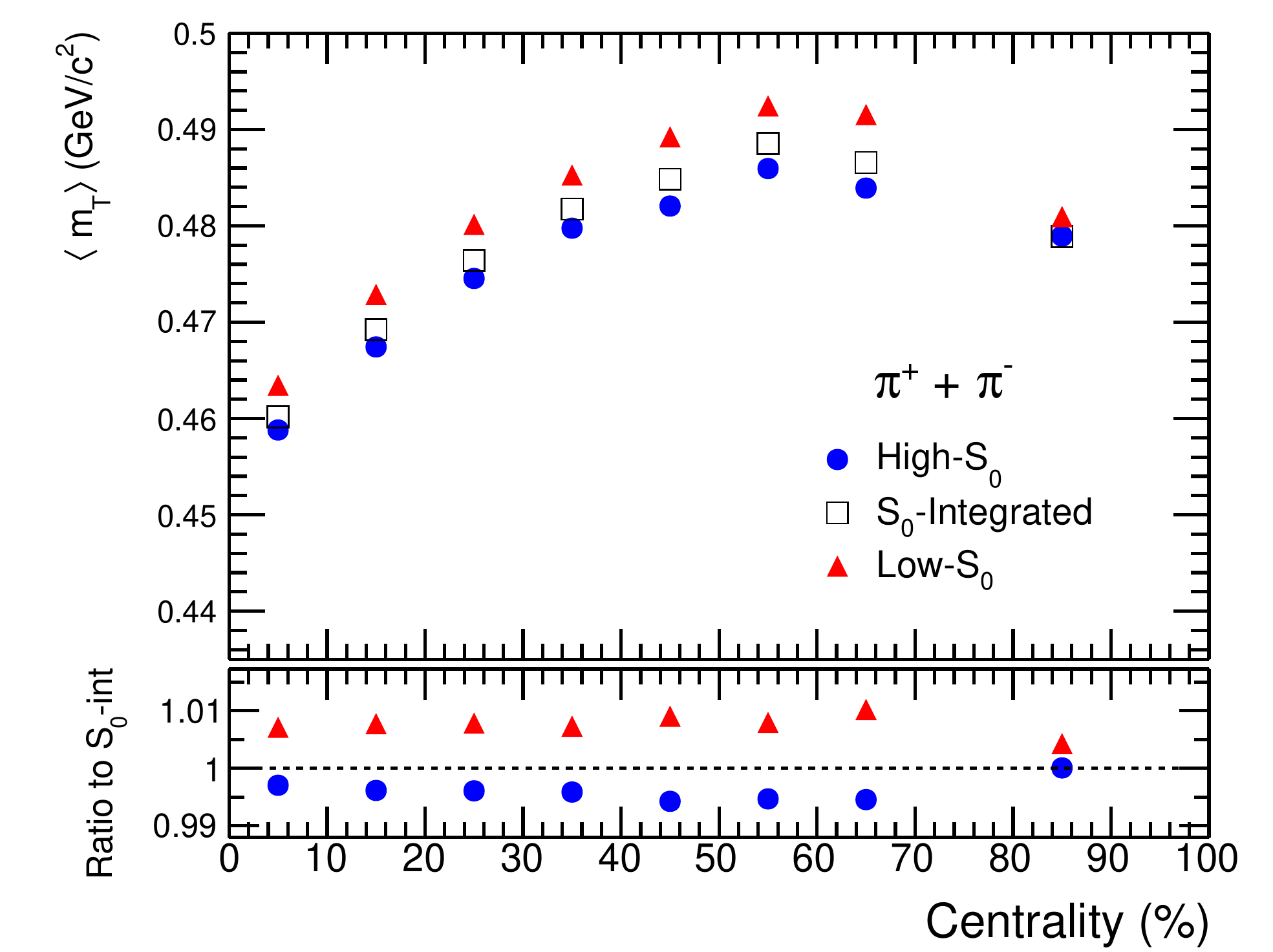}
\includegraphics[scale=0.44]{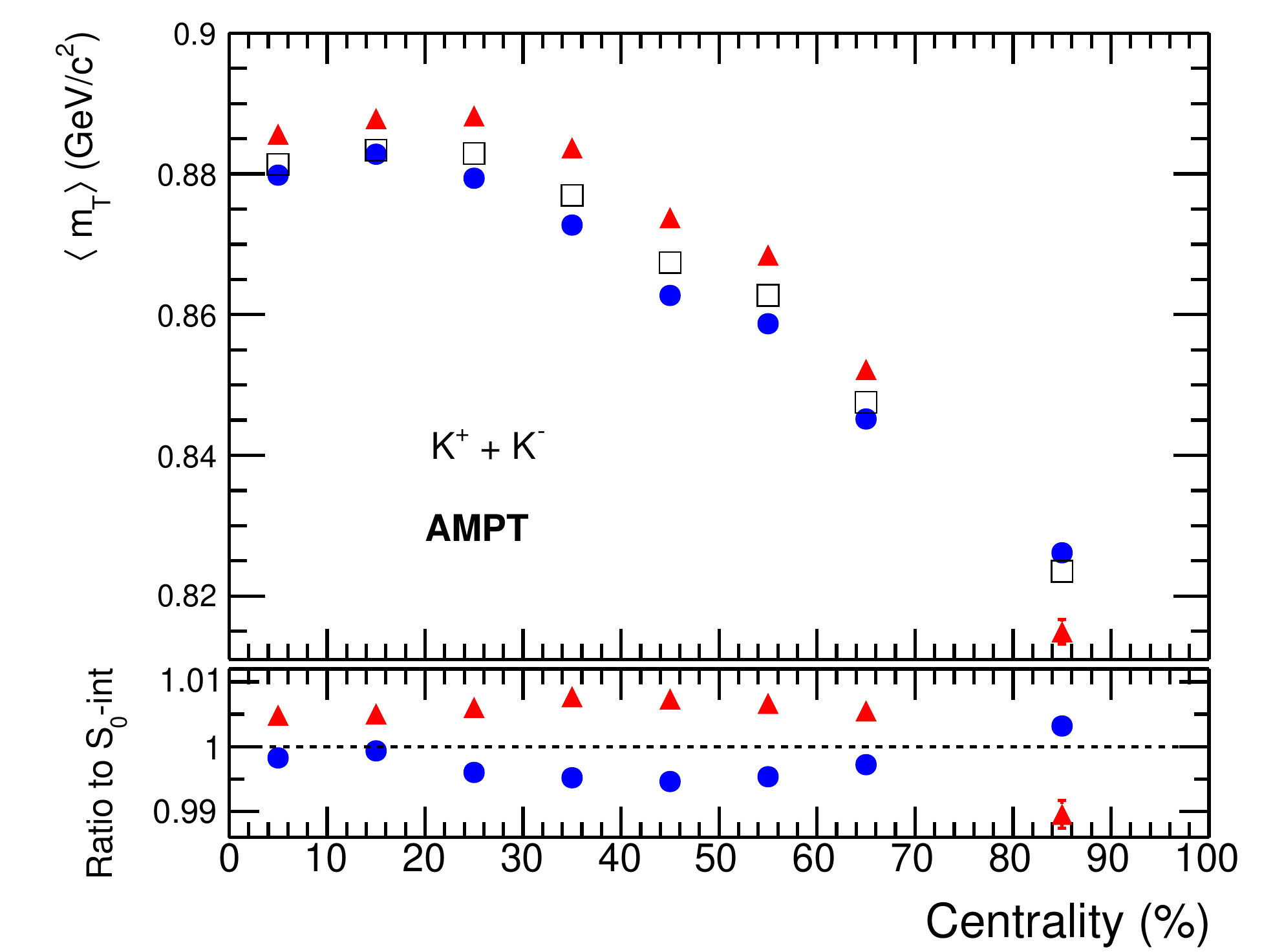}
\includegraphics[scale=0.44]{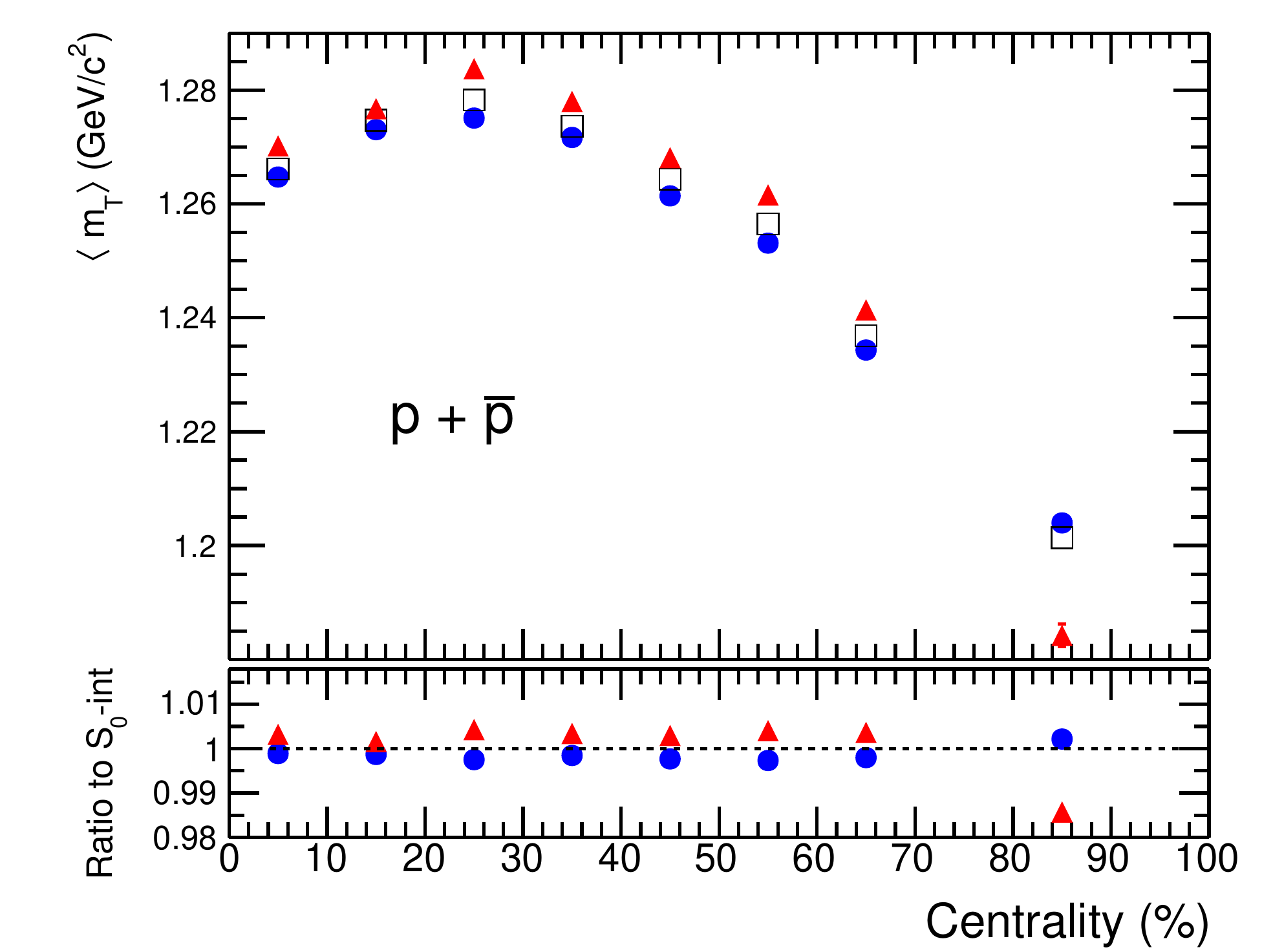}
\caption[width=18cm]{(Color Online) $\langle m_{\rm T}\rangle$ vs. collision centrality(\%) for pions (top), kaons (middle) and protons (bottom) in Pb-Pb collisions with high-$\rm S_{0}$, $\rm S_0$-integrated and low-$\rm S_{0}$ events, respectively }
\label{mean_mT}
\end{center}
\end{figure}

\begin{figure}[ht!]
\begin{center}
\includegraphics[scale=0.4]{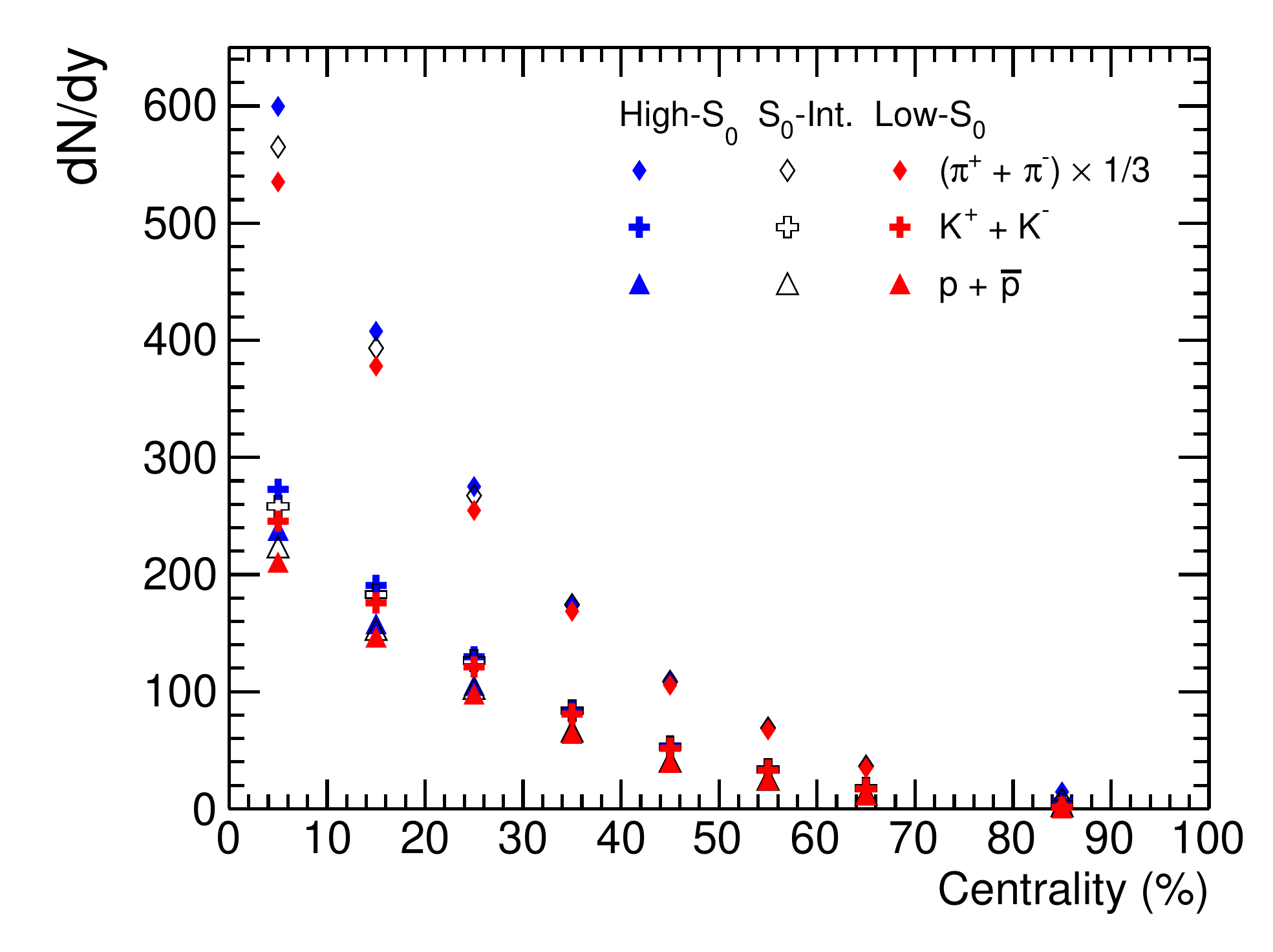}
\caption[width=18cm]{(Color Online) Integrated yield (dN/dy) vs. collision centrality (\%) for pions (top), kaons (middle) and protons (bottom) at mid-rapidity in Pb-Pb collisions.}
\label{intYield}
\end{center}
\end{figure}

\begin{figure}[ht!]
\begin{center}
\includegraphics[scale=0.4]{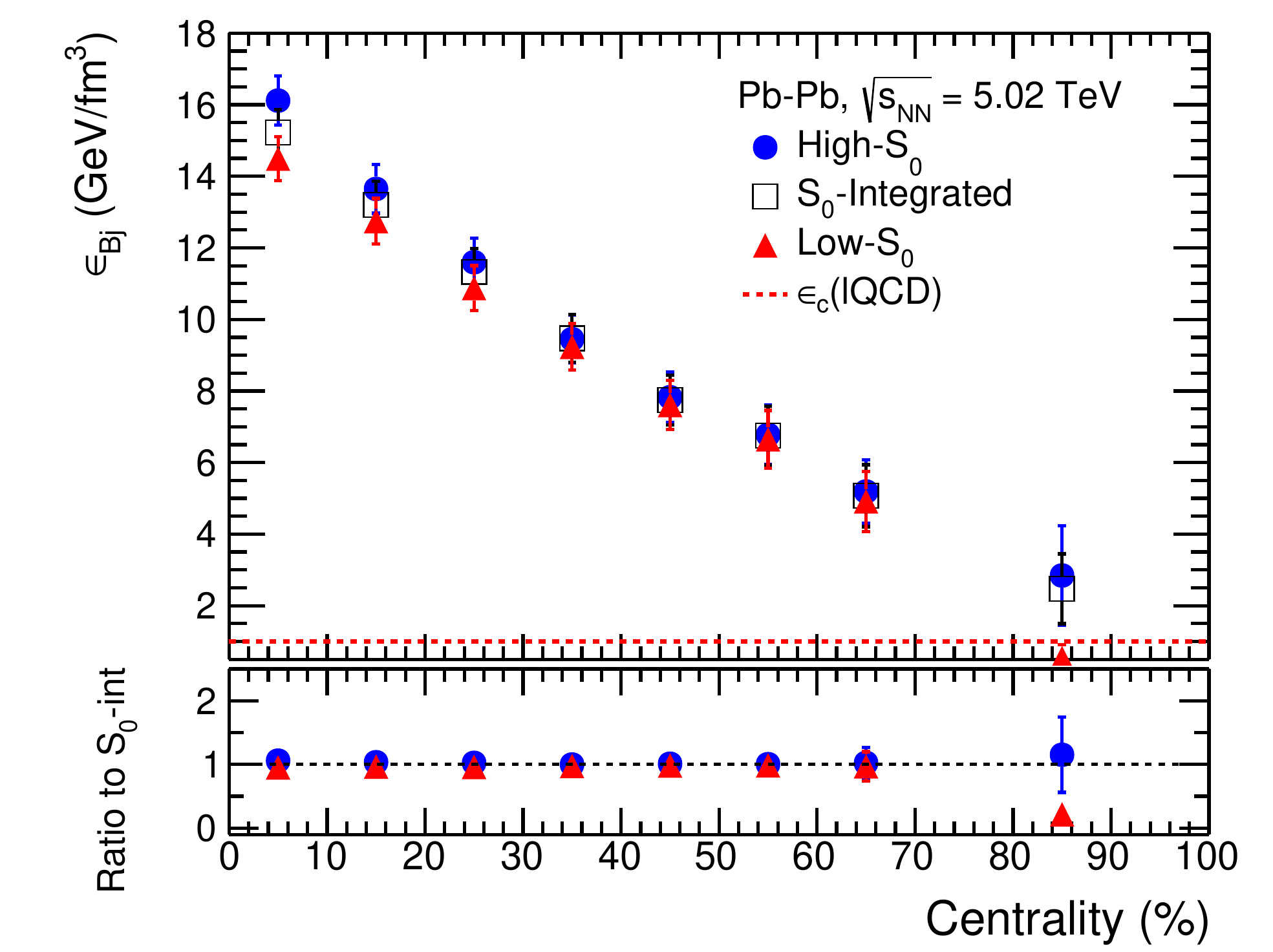}
\caption[width=18cm]{(Color Online) Top plot: Bjorken energy density~($\epsilon_{\rm Bj}$) vs. centrality~(\%) with high-$\rm S_{0}$, $\rm S_0$-integrated and low-$\rm S_{0}$ events in Pb-Pb collisions. Bottom plot: ratio of Bjorken energy density~($\epsilon_{\rm Bj}$) for high-$\rm S_{0}$ and low-$\rm S_{0}$ events to the $\rm S_0$-integrated events. The dotted line shows lattice QCD predicted value of critical energy density for a deconfinement transition.}
\label{BjorkenEnergyDensity}
\end{center}
\end{figure}

In heavy-ion collisions, the transverse energy ($E_{\rm T}$) is one of the significant global observables that is used to study the possible formation of a medium of quarks and gluons under extreme temperature and energy density. Before the collisions, all the energy is carried by the beam particles in longitudinal phase space. But after the collisions, the final state particle production in the transverse plane carries finite transverse energy ($E_{\rm T}$), which is an event-by-event observable and it is closely related to the collision geometry.  In the Bjorken boost-invariant hydrodynamics model~\cite{Bjorken:1982qr} for relativistic heavy-ion collisions, $E_{\rm T}$ at mid-rapidity gives the quantitative estimation of the initial energy density produced in an interaction. Under boost invariance, the Bjorken energy density ($\epsilon_{\rm Bj}$) in the nuclear collision zone can be estimated as,

\begin{eqnarray}
\epsilon_{\rm Bj} = \frac{1}{\tau S_{\rm T}}\frac{dE_{\rm T}}{dy} .
\label{E1}
\end{eqnarray}

where, $\tau$ is the formation time and usually taken to be $1~{\rm fm/c}$. $E_{\rm T}$ is the total transverse energy and $S_{\rm T} = \pi R^2$ is the transverse overlap area of the colliding nuclei. As $R = R_{0} A^{1/3}$, one replaces $A = N_{part}/2$. That makes the expression for transverse overlap area,
\begin{eqnarray}
S_{\rm T} = \pi R_{0}^2 \left( \frac{N_{part}}{2}\right)^{2/3}.
\label{TransArea}
\end{eqnarray}

Transverse energy ($E_{\rm T}$) at mid-rapidity region can be approximated as~\cite{ALICE:2016igk,Sahoo:2014aca,STAR:2008med},
\begin{equation}
\begin{gathered}
\frac{dE_{\rm T}}{dy} \approx  \frac{3}{2}\times\left(\langle m_{\rm T} \rangle \frac{dN}{dy}\right)_{\pi^{\pm}} + 2\times\left(\langle m_{\rm T} \rangle\frac{dN}{dy}\right)_{K^{\pm},p,\bar{p}}.
\label{E2}
\end{gathered}
\end{equation}
The multiplicative factors 3/2 and 2 account for neutral particles. $ m_{\rm T} = \sqrt{p_{\rm T}^2 + m^2}$, is the transverse mass and $dN/dy$ is the integrated yield for $\pi^{\pm}$, $K^{\pm}$ and $p+\bar{p}$ at mid-rapidity region {\em i.e.} $|y| < 0.5 $, estimated for $p_{\rm{T}}>$~0.15~GeV/$c$.

Figure \ref{mean_mT} shows the mean transverse mass ($\langle m_{\rm T} \rangle$) as a function of different centrality classes in Pb-Pb collisions at $\sqrt{s_{\rm NN}}=5.02 ~\rm{TeV}$ at mid-rapidity for $\pi^{\pm}$, $K^{\pm}$ and $p+\bar{p}$ using AMPT model. Except
pions, where resonance decay contributions are expected, for all other charged particles $\langle m_{\rm T} \rangle$ shows a
decrease towards peripheral collisions. The lower panels of the figure show the effect of event topology on $\langle m_{\rm T} \rangle$, 
where one observes higher $\langle m_{\rm T} \rangle$ for low-$\rm S_0$ events (jetty).
Figure \ref{intYield} shows the integrated yield ($dN/dy$) as a function of different centrality classes in Pb-Pb collisions at $\sqrt{s_{\rm NN}}=5.02 ~\rm{TeV}$ at mid-rapidity for different identified particles using AMPT model. As expected, the integrated yield is higher for more central collisions and gradually decreases as we move towards mid-central and peripheral collisions. It can be accounted due to the decrease in the participating partonic matter from central to peripheral collisions. For pions, the integrated yield is higher than kaon and proton, which follows a thermalised Boltzmannian production of particles in a multiparticle production process. The integrated yield as a function of spherocity shows that high-$\rm S_{0}$ events have higher yield than low-$\rm S_{0}$ events. It is also clear that the integrated  yield  is highly  dependent  on  the  spherocity classes for most central heavy-ion collisions and the dependence  decreases  while  going  towards  peripheral  collisions. 

Figure~\ref{BjorkenEnergyDensity} shows the  Bjorken energy density~($\epsilon_{\rm Bj}$) vs. centrality~(\%) for high-$\rm S_{0}$, $\rm S_0$-integrated and low-$\rm S_{0}$ events in Pb-Pb collisions. We observe a strong dependence of Bjorken energy density on the centrality classes. The values of the initial energy density is observed to be higher than the lattice QCD estimation of 1 GeV/fm$^3$ energy density for a deconfinement transition~\cite{Karsch:2001vs}. However, we found that the Bjorken energy density is independent of the spherocity selection and irrespective of the event topology, it is similar for both high-$\rm S_{0}$ and low-$\rm S_{0}$ events for all collision centralities. It is noteworthy that the Bjorken energy density for a given collision centrality has 
effects from the number of particles produced and their mean transverse mass. Because of the opposite trends of these two observables
 with event spherocity, the overall effect cancels out leaving Bjorken energy density to be independent of event topology.

\subsection{Squared speed of sound ($c_{\rm s}^2$) and pseudorapidity distribution}

\begin{figure}[ht!]
\begin{center}
\includegraphics[scale=0.44]{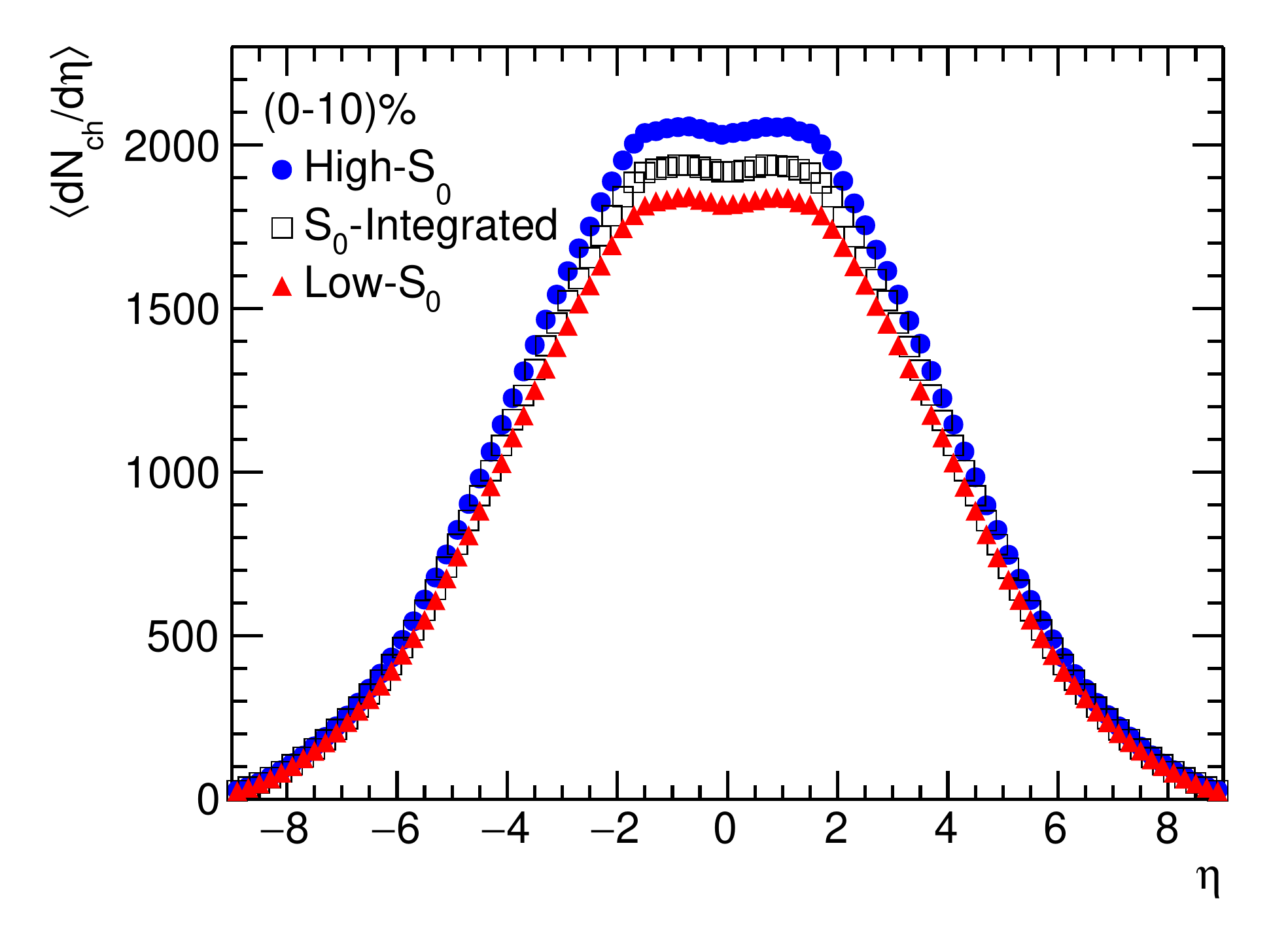}
\includegraphics[scale=0.44]{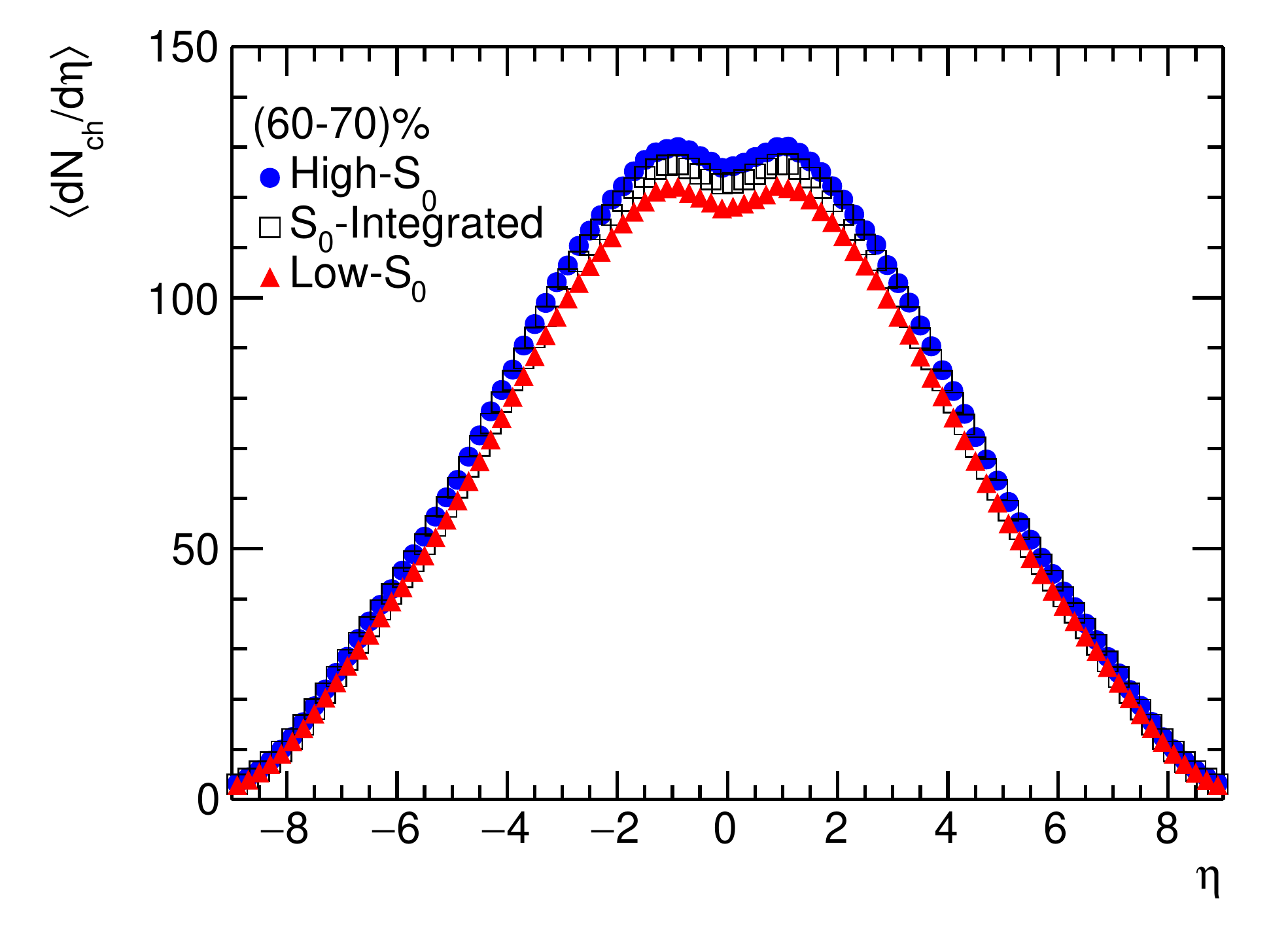}
\caption[width=12cm]{(Color Online) Charged particles pseudorapidity distribution ($dN_{ch}/d{\eta}$) for high-$\rm S_{0}$, $\rm S_0$-integrated and low-$\rm S_{0}$ events for (0-10)\% (top) and (60-70)\% (bottom) centrality class in Pb-Pb collisions at $\sqrt{s_{\rm NN}}$ = 5.02 TeV.}
\label{pseudorapidity}
\end{center}
\end{figure}

\begin{figure}[ht!]
\begin{center}
\includegraphics[scale=0.4]{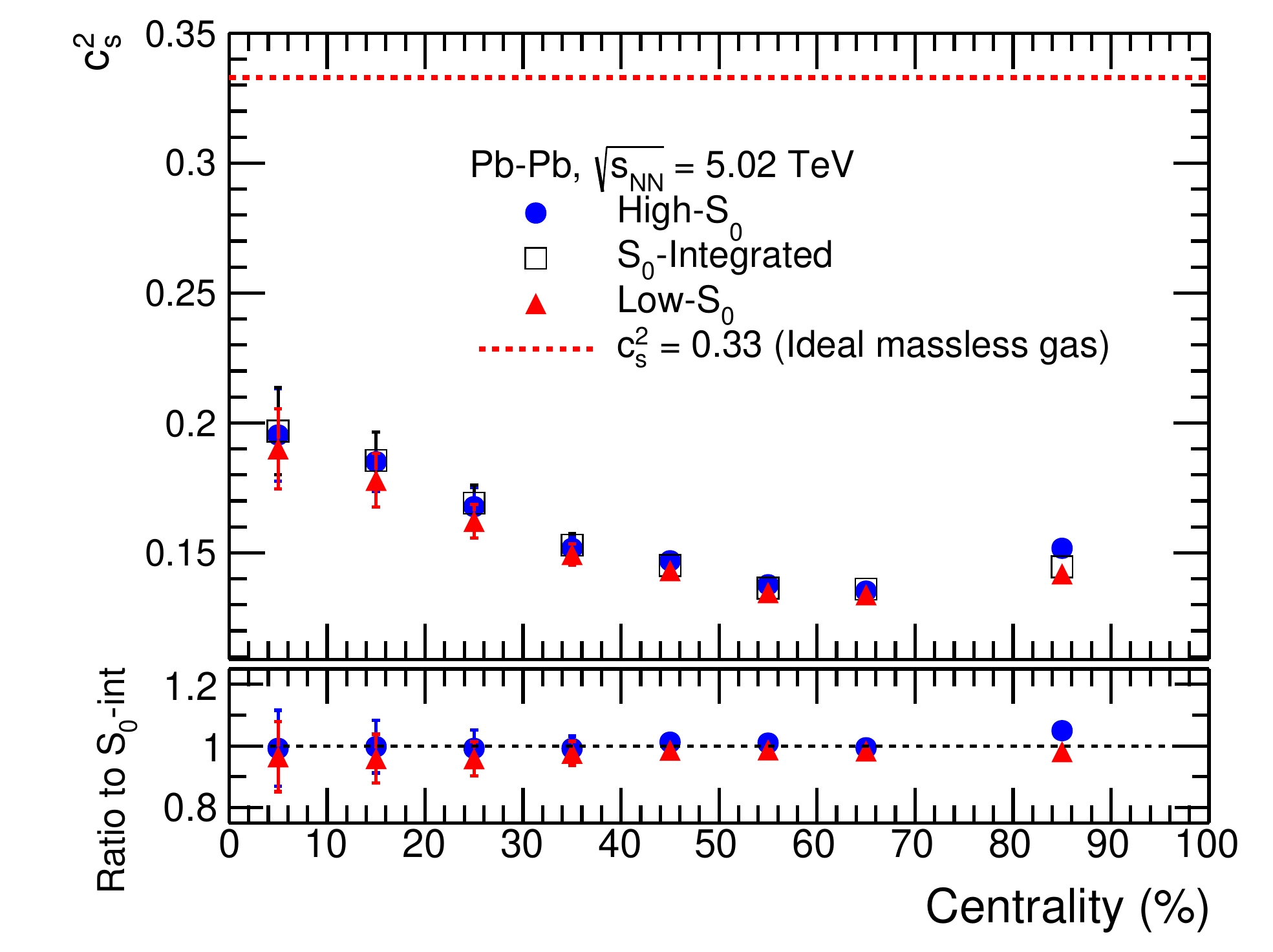}
\caption[width=18cm]{(Color Online) Top plot: Speed of sound ($c_{\rm s}^2$) vs. Centrality(\%) for high-$\rm S_{0}$, $\rm S_0$-integrated and low-$\rm S_{0}$ events in Pb-Pb collisions. Bottom plot: Ratio of speed of sound ($c_{\rm s}^2$) for high-$\rm S_{0}$ and low-$\rm S_{0}$ events to the $S_0$-integrated events.}
\label{cs2plot}
\end{center}
\end{figure}

Figure~\ref{pseudorapidity} shows pseudorapidity distributions of charged particles in Pb-Pb collisions at $\sqrt{s_{\rm{NN}}}$ = 5.02 TeV at mid rapidity for (0-10)\% and (60-70)\% centrality classes in different spherocity classes. Figure~\ref{pseudorapidity} complements the observation seen in Fig.~\ref{intYield}, where the charged particle multiplicity is found to be higher for high-$S_{0}$ events compared to low-$S_{0}$ events.

In Landau hydrodynamical model~\cite{Landau:1953gs}, the width of the rapidity distribution is related to the speed of sound ($c_s$) via the following expression.
\begin{equation}
\sigma_{y}^2 = \frac{8}{3}\frac{c_s^2}{1-c_s^2}ln\bigg(\frac{\sqrt{s_{\rm{NN}}}}{2m_p}\bigg).
\label{eq-cs2}
\end{equation}
Here, $m_p$ is the mass of proton and $\sigma_{y}$ is the width of the rapidity distribution and $c_s^2 = 1/3$ for ideal gas. Due to presence of a dip, it is difficult to fit a single Gaussian function to the pseudorapidity distribution. Generally, in experiments~\cite{ALICE:2013jfw} a double Gaussian function is used to fit the pseudorapidity distribution, which is given by,
\begin{equation}
A_{1}e^{\frac{-\eta^2}{2\sigma_1^{2}}} - A_{2}e^{\frac{-\eta^2}{2\sigma_2^{2}}}.
\label{eq-doublegaus}
\end{equation}

Here, $A_{1}$ and $A_{2}$ are normalisation parameters and $\sigma_{1}$ and $\sigma_{2}$ are the widths of the double Gaussian distribution. After fitting  we have obtained the values of  $\sigma_{1}$ and $\sigma_{2}$. The values of $\sigma_{1}$ and $\sigma_{2}$ are given in Table~\ref{tab:sigma}. The fitting has been performed using $\chi^{2}$ minimisation method and corresponding $\chi^{2}/\text{ndf}$ values for the fittings for each spherocity classes across all centralities are shown in Table~\ref{tab:chi2byndf}. Thus, $\sigma_{1}$ has been used as the default value and the maximum deviation of $\sigma_{1}$ and $\sigma_{2}$ is used for the uncertainty calculation of $c^2_{s}$. Figure~\ref{cs2plot} shows the squared speed of sound as a function of centrality in Pb-Pb at $\sqrt{s_{\rm{NN}}}$ = 5.02 TeV for different spherocity classes. The $c^2_{s}$ value as a function of centrality shows that for the central collision system, it is higher and gradually decreases toward peripheral collisions. It indicates that central collisions are denser compared to the peripheral collisions. However, $c^2_{s}$ is found to be similar for all spherocity classes within uncertainty.

\begin{table*} [!hpt]
                \centering
                \scalebox{1}{
                \begin{tabular}{|c |c |c | c| c | c| c|}
                \hline
                \multirow{2}{*}{Centrality(\%)} & \multicolumn{2}{c|}{{\textbf{High-$S_{0}$}}} & \multicolumn{2}{c|}{{\textbf{$S_{0}$ Integrated}}} & \multicolumn{2}{c|}{{\textbf{ Low-$S_{0}$}}} \\ \cline{2-7} 
                & $\sigma_{1}$ & $\sigma_{2}$ & $\sigma_{1}$ & $\sigma_{2}$ & $\sigma_{1}$ & $\sigma_{2}$ \\
                \hline
             	\hline
                0--10            &2.110 $\pm$ 0.024      &1.757 $\pm$ 0.223       &2.120 $\pm$ 0.023    &1.786 $\pm$ 0.022     &2.079 $\pm$ 0.022      &1.769 $\pm$ 0.020 \\
                \hline 
                10--20           &2.050 $\pm$ 0.019      &1.819 $\pm$ 0.018       &2.053 $\pm$ 0.016    &1.830 $\pm$ 0.016     &2.007 $\pm$ 0.017      &1.797 $\pm$ 0.016 \\
                \hline
                20--30           &1.946 $\pm$ 0.011      &1.794 $\pm$ 0.019       &1.954 $\pm$ 0.010    &1.807 $\pm$ 0.010     &1.911 $\pm$ 0.010      &1.773 $\pm$ 0.016 \\
                \hline
                30--40           &1.845 $\pm$ 0.006      &1.751 $\pm$ 0.006       &1.853 $\pm$ 0.006    &1.759 $\pm$ 0.006     &1.830 $\pm$ 0.006      &1.740 $\pm$ 0.006 \\
                \hline
                40--50           &1.815 $\pm$ 0.004      &1.747 $\pm$ 0.004       &1.803 $\pm$ 0.004    &1.745 $\pm$ 0.004     &1.790 $\pm$ 0.004      &1.731 $\pm$ 0.004 \\
                \hline
                50--60           &1.755 $\pm$ 0.001      &1.742 $\pm$ 0.001       &1.746 $\pm$ 0.001    &1.733 $\pm$ 0.001     &1.735 $\pm$ 0.003      &1.722 $\pm$ 0.003 \\
                \hline
                60--70           &1.738 $\pm$ 0.034      &1.731 $\pm$ 0.034       &1.743 $\pm$ 0.003    &1.737 $\pm$ 0.003     &1.729 $\pm$ 0.003      &1.722 $\pm$ 0.003 \\
                \hline
                70--100          &1.846 $\pm$ 0.001      &1.843 $\pm$ 0.001       &1.799 $\pm$ 0.001    &1.798 $\pm$ 0.001     &1.782 $\pm$ 0.001      &1.781 $\pm$ 0.001 \\
               
                \hline
                \end{tabular} 
                }              
                \caption{ Double Gaussian width parameters from fitting the pseudorapidity distributions in the range $|\eta|<2$ using Eq.~\ref{eq-doublegaus}. \label{tab:sigma}}
\end{table*}

\begin{table*}[ht!]
 \centering
\begin{tabular}{|l|l|l|l|}
\hline
\multirow{2}{*}{Centrality(\%)} & \multicolumn{3}{l|}{$\chi^2$/ndf}       \\ \cline{2-4} 
                                & High-$\rm S_0$ & $\rm S_0$ Integrated & Low-$\rm S_0$    \\ \hline

0-10                           & 0.16   & 0.11       & 0.13 \\ \hline
10-20                          & 0.55   & 0.45       & 0.30 \\ \hline
20-30                          & 0.92   & 0.99       & 1.15 \\ \hline
30-40                          & 1.16   & 1.28       & 0.97 \\ \hline
40-50                          & 1.06   & 1.17       & 0.92 \\ \hline
50-60                          & 0.64   & 0.86       & 0.96 \\ \hline
60-70                          & 0.67   & 0.64       & 0.91 \\ \hline
70-100                         & 0.51   & 0.35       & 1.15 \\ \hline

\end{tabular}
\caption[width=18cm]{ $\chi^2$ /ndf values for the fitting of $dN_{ch}/d{\eta}$ to a double Gaussian distribution.}
\label{tab:chi2byndf}
\end{table*}

\subsection{Kinetic Freeze-out Properties}

\begin{figure*}[ht!]
\begin{center}
\includegraphics[scale=0.29]{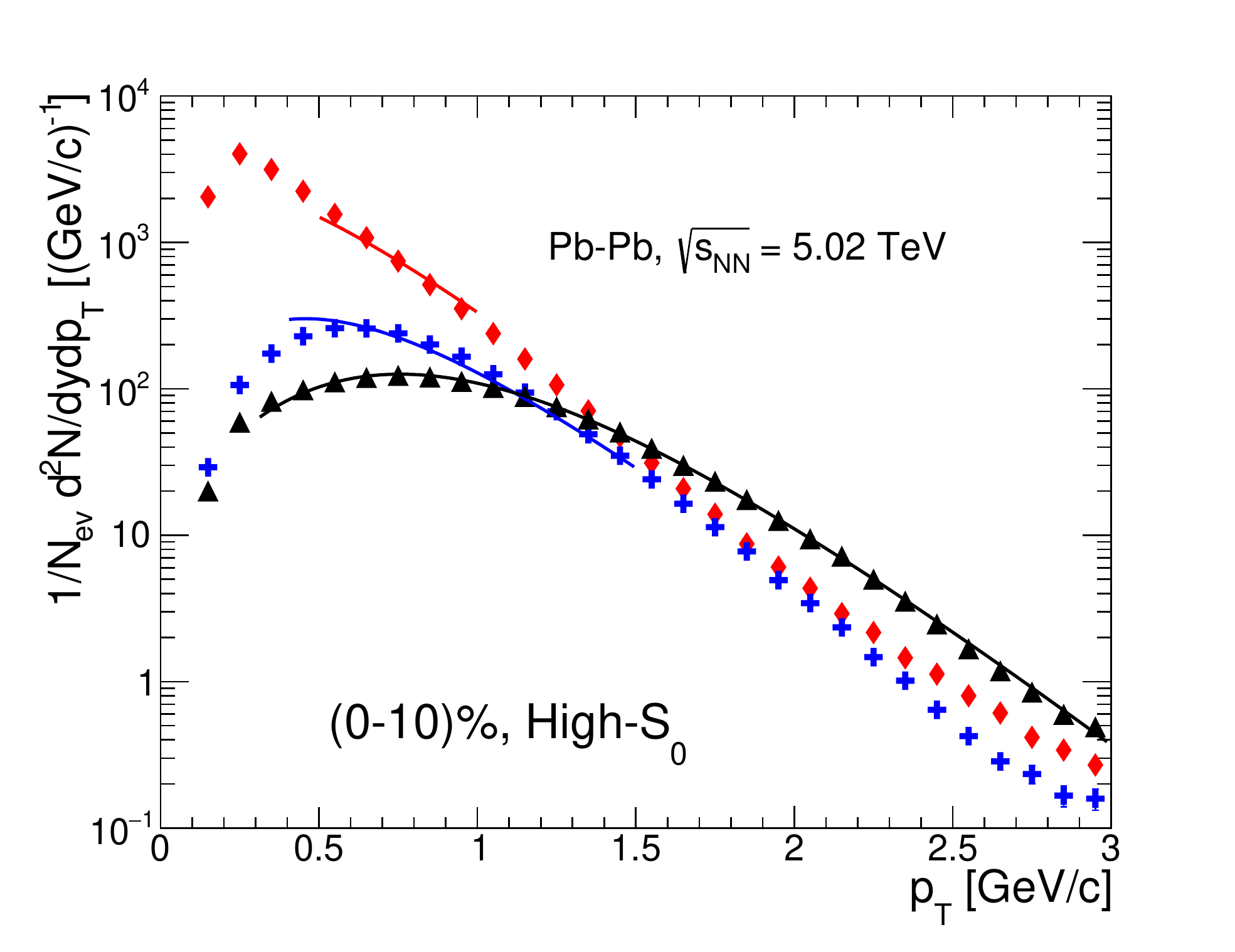}
\includegraphics[scale=0.29]{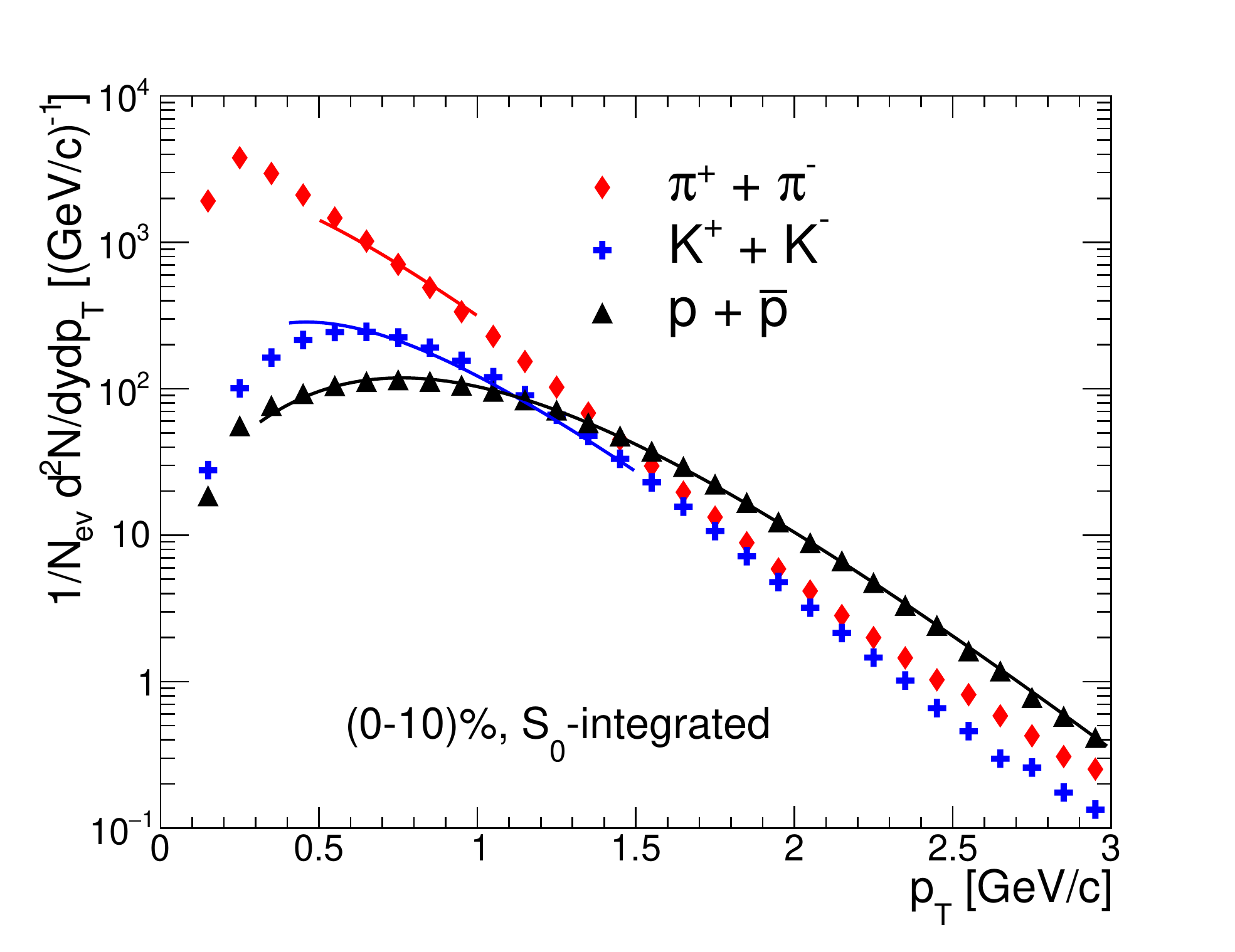}
\includegraphics[scale=0.29]{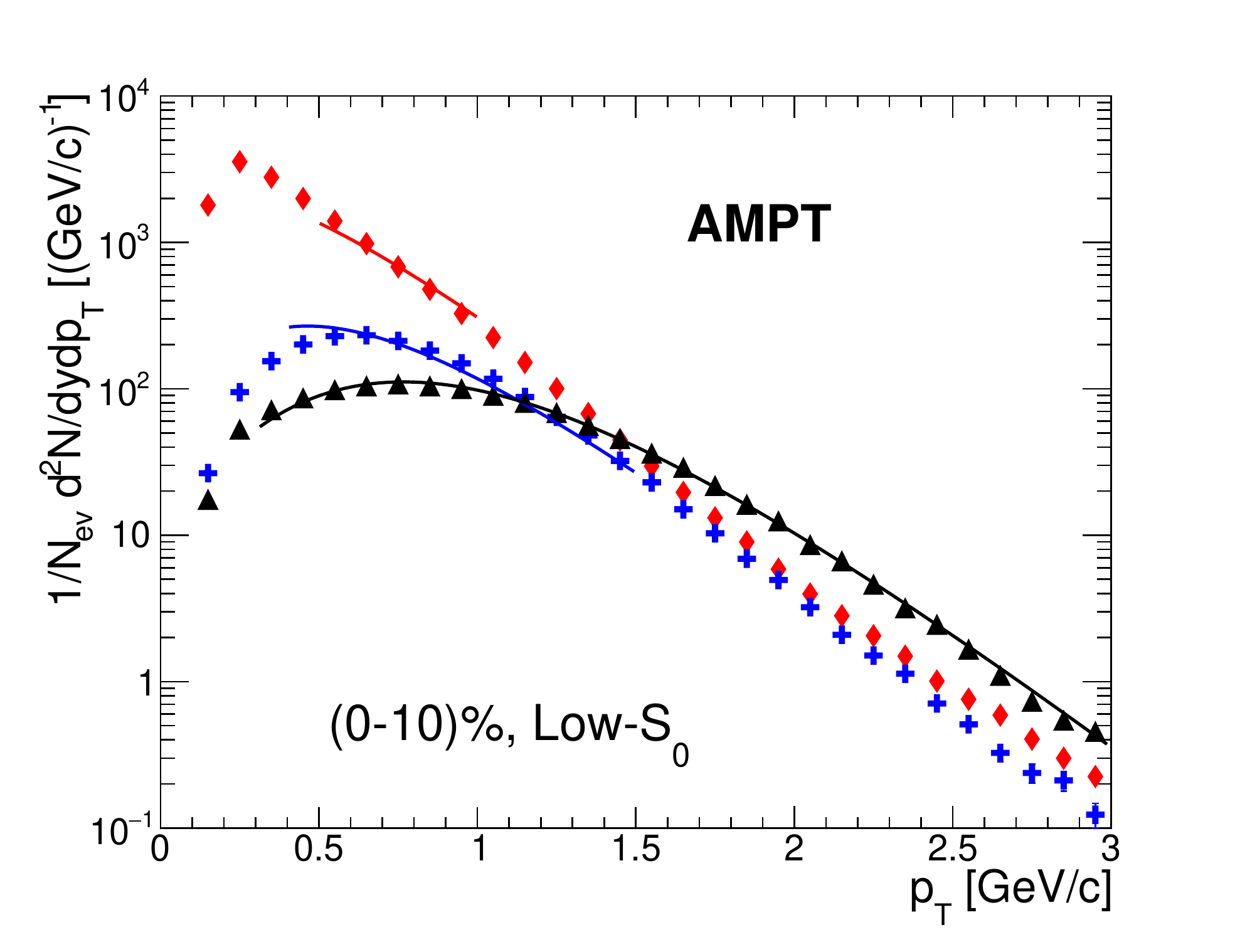}
\includegraphics[scale=0.29]{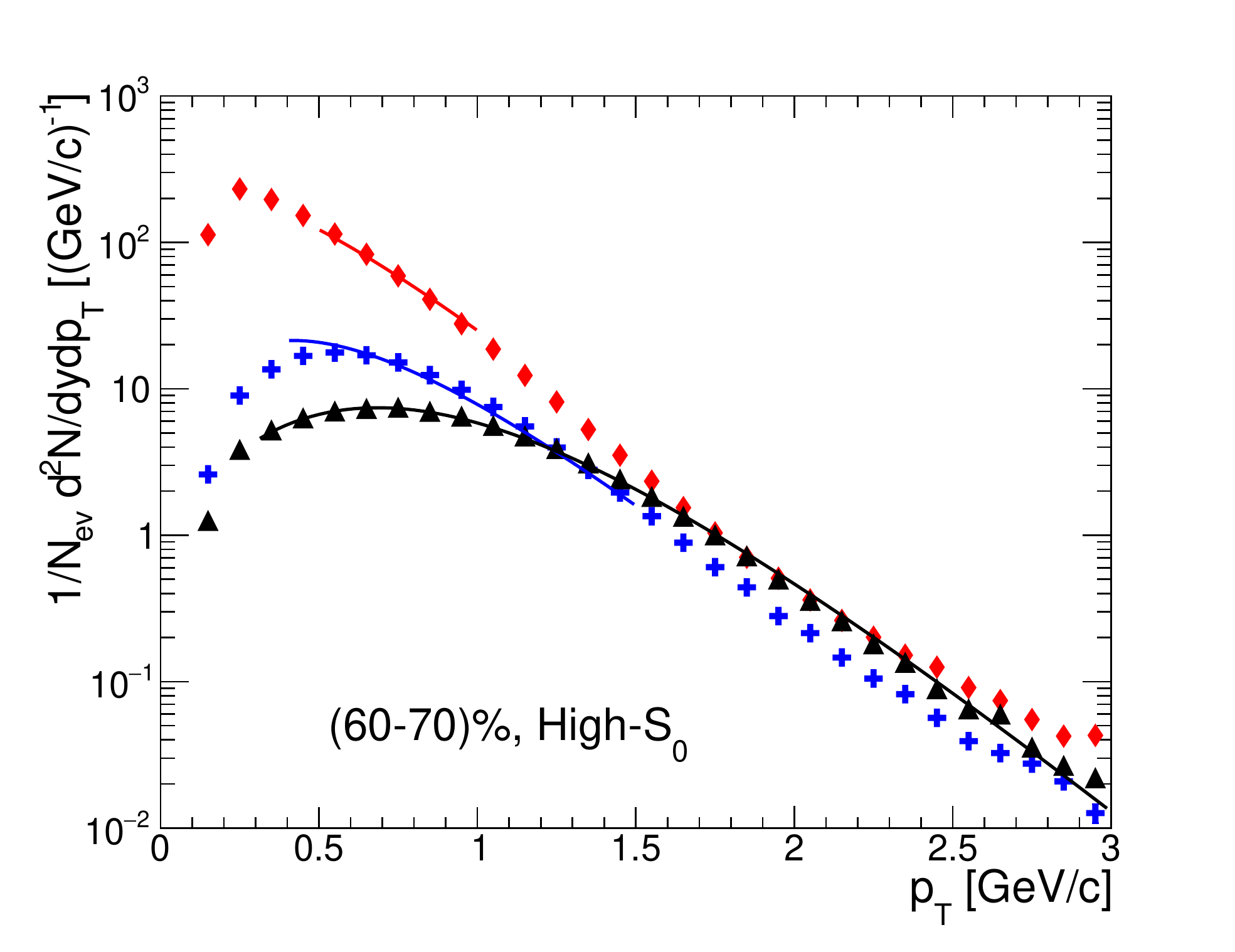}
\includegraphics[scale=0.29]{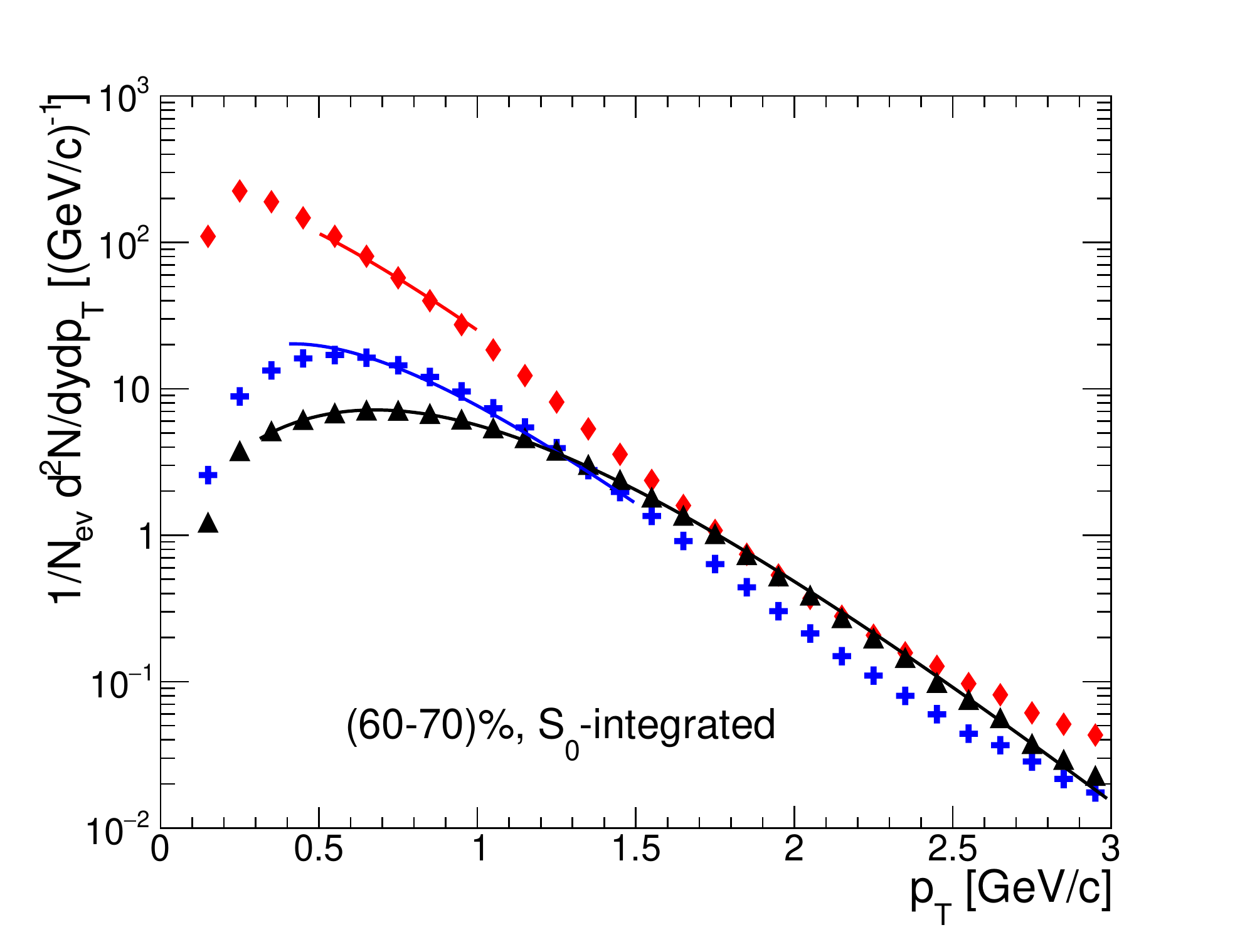}
\includegraphics[scale=0.29]{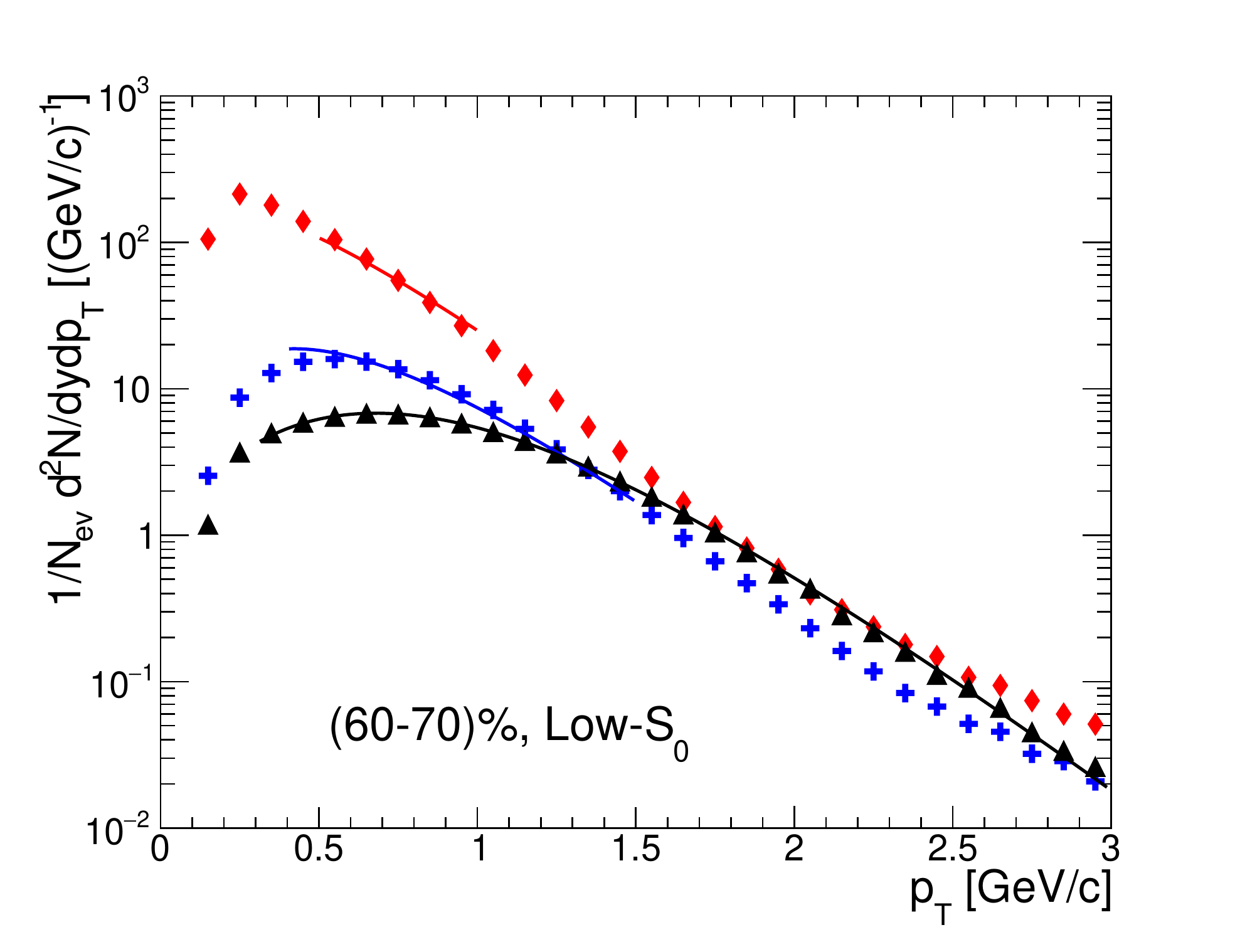}
\caption[width=18cm]{(Color Online) Simultaneous fitting of identified charged particles' $p_{\rm{T}}$ spectra with Boltzmann-Gibbs blastwave function for high-$\rm{S_{0}}$ (left), $\rm{S_{0}}$-integrated (middle) and low-$\rm{S_{0}}$ (right) events in Pb-Pb collisions at $\rm{\sqrt{s_{NN}}}$ = 5.02 TeV. The fitting is shown for (0-10)$\%$ (top) and (60-70)$\%$ (bottom) centrality classes.}
\label{BGBWfitplot}
\end{center}
\end{figure*}

The hot and dense medium formed in relativistic heavy-ion collisions cools down as the system expands until the kinetic freeze-out is achieved. At kinetic freeze-out, the transverse momentum spectra of the particles are frozen, which carries information about the phase-space distribution of the final state of the fireball produced during relativistic heavy-ion collisions. The Boltzmann-Gibbs blastwave (BGBW) distribution~\cite{Schnedermann:1993ws} can be used to describe the transverse momentum spectra of identified charged particles formed in heavy-ion collisions and one can obtain the transverse radial flow velocity ($\rm{\beta_{T}}$) and kinetic freeze-out temperature ($\rm{T_{kin}}$) of the system. The invariant yield in the BGBW framework can be expressed as:
 \begin{eqnarray}
\label{bgbw1}
E\frac{d^3N}{dp^3}=C \int d^3\sigma_{\mu} p^{\mu} exp(-\frac{p^{\mu} u_{\mu}}{T_{\rm kin}}).
\end{eqnarray}
Here, $C$ is the normalisation constant and $p^{\mu}$, the particle four momentum is given by, 
 \begin{eqnarray}
p^{\mu}~=~(m_{T}{\cosh}y,~p_{T}\cos\phi,~ p_{T}\sin\phi,~ m_{T}{\sinh}y).
\end{eqnarray}
The particle four-velocity is given by,
 \begin{eqnarray}
u^{\mu}=\cosh\rho~(\cosh\eta,~\tanh\rho~\cos\phi_{r},~\tanh\rho~\sin
\phi_{r},~\sinh~\eta),
\end{eqnarray}
so that
 \begin{eqnarray}
p^{\mu}u_{\mu} = m_T \cosh (y-\eta) \cosh \rho -p_T \sinh \rho \cos(\phi-\phi_r). 
\end{eqnarray}

The freeze-out surface is parametrised as, 
 \begin{eqnarray}
d^3\sigma_{\mu}~=~(\cosh\eta,~0,~0, -\sinh\eta)~\tau~r~dr~d\eta~d\phi_{r}.
\end{eqnarray}
Here, $\eta$ is the space-time rapidity. Now, Eq.~\ref{bgbw1} can be written as,
 \begin{eqnarray}
\label{boltz_blast}
\left.\frac{d^2N}{dp_{T}dy}\right|_{y=0} = C~ p_T~m_T \int_{0}^{R_{0}} r\;dr\;K_{1}\Big(\frac{m_{T}\;\cosh\rho}{T_{\rm kin}}\Big)I_{0}
\Big(\frac{p_{T}\;\sinh\rho}{T_{\rm kin}}\Big).
\end{eqnarray}
 Here, $K_{1}\displaystyle\Big(\frac{m_{T}\;{\cosh}\rho}{T_{\rm kin}}\Big)$ and $I_{0}\displaystyle\Big(\frac{p_{T}\;{\sinh}\rho}{T_{\rm kin}}\Big)$ are modified Bessel's functions, which are given by,
\begin{eqnarray}
\centering
K_{1}\Big(\frac{m_{T}\;{\cosh}\rho}{T_{\rm kin}}\Big)=\int_{0}^{\infty} {\cosh}y\;{\exp} \Big(-\frac{m_{T}\;{\cosh}y\;{\cosh}\rho}{{T_{\rm kin}}}\Big)dy\nonumber,
\end{eqnarray}
\begin{eqnarray}
\centering
I_{0}\Big(\frac{p_{T}\;{\sinh}\rho}{T_{\rm kin}}\Big)=\frac{1}{2\pi}\int_0^{2\pi} \exp\Big(\frac{p_{T}\;{\sinh}\rho\;{\cos}\phi}{{T_{\rm kin}}}\Big)d\phi \nonumber,
\end{eqnarray}
where $\rho={\tanh}^{-1}\beta_{\rm T}$ and $\rm{\beta_{T}=\displaystyle\beta_s\xi^n}$ \cite{Schnedermann:1993ws,Huovinen:2001cy,BraunMunzinger:1994xr, Tang:2011xq}. $\rm{\beta_{T}}$ is called radial flow velocity, $\xi=(r/R_0)$, $\rm{\beta_s}$ is the maximum surface velocity, $r$ is the radial distance and $R_0$ is the maximum radius of the source at freeze-out. In this model, the particles closer to the center of the fireball are assumed to move slower than the ones at the edges. The mean transverse velocity is given by \cite{Adcox:2003nr}, 
 \begin{eqnarray}
\langle\beta_{\rm T}\rangle =\frac{\int \beta_s\xi^n\xi\;d\xi}{\int \xi\;d\xi}=\Big(\frac{2}{2+n}\Big)\beta_s .
\end{eqnarray}
Figure \ref{BGBWfitplot} shows the simultaneous BGBW fitting to the identified charged particles' $p_{\rm{T}}$ spectra in (0-10)\% and (60-70)\% centrality classes. The fitting ranges for pions, kaons and protons are (0.5 - 1) GeV/c, (0.4 - 1.5) GeV/c and (0.3 - 3) GeV/c, respectively. The fitting has been performed using $\chi^{2}$ minimisation method keeping 
$\rm{T_{kin}}$, $\rm{\beta_{s}}$ and $n$ as free parameters. The corresponding $\chi^{2}/\text{ndf}$ values for the fittings for each spherocity classes across all centralities are shown in Table~\ref{tab:BGBWfittable}.

\begin{table*}[ht!]
  \centering
\begin{tabular}{|l|l|l|l|}
\hline
\multirow{2}{*}{Centrality(\%)} & \multicolumn{3}{l|}{$\chi^2$/ndf}       \\ \cline{2-4} 
                                & High-$\rm S_0$ & $\rm S_0$ Integrated & Low-$\rm S_0$    \\ \hline
0-10                           & 2.7   & 2.8       & 3.2 \\ \hline
10-20                           & 2.5   & 2.8       & 2.8 \\ \hline
20-30                           & 2.9   & 2.8       & 2.7 \\ \hline
30-40                           & 2.8   & 2.6       & 2.6 \\ \hline
40-50                           & 2.4   & 2.2       & 2.1 \\ \hline
50-60                           & 2.1   & 2.1       & 2.0 \\ \hline
60-70                           & 2.4   & 2.2       & 1.6 \\ \hline
70-100                           & 2.2   & 2.1       & 2.4 \\ \hline

\end{tabular}
\caption[width=18cm]{ $\chi^2$/ndf values for the simultaneous fitting of identified charged particles' $p_{\rm{T}}$ spectra to the BGBW distribution.}
\label{tab:BGBWfittable}
\end{table*}

\begin{figure}[ht!]
\begin{center}
\includegraphics[scale=0.43]{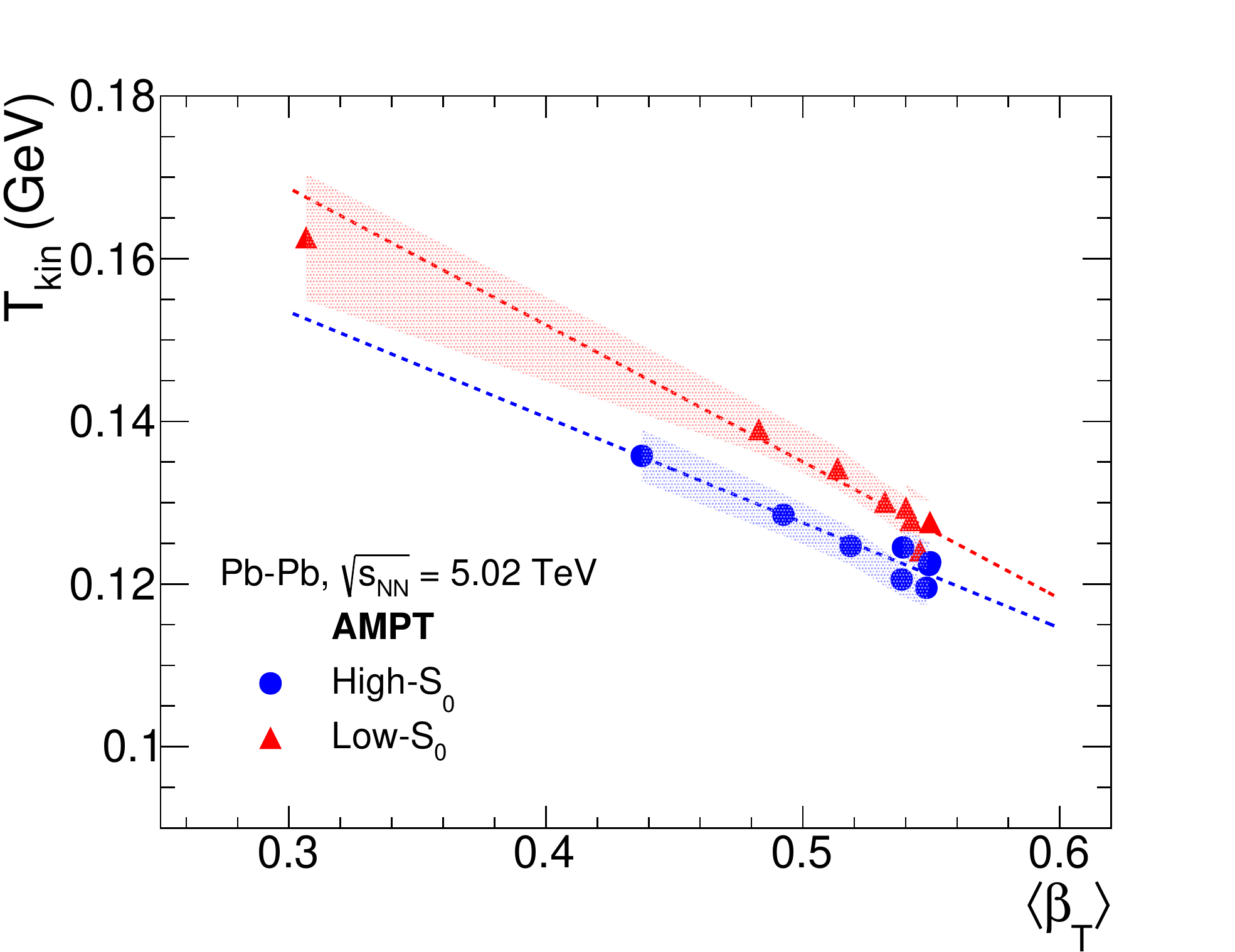}
\caption{(Color Online) Kinetic freeze-out temperature versus transverse radial flow obtained from simultaneous fit of identified particles’ $p_{\rm{T}}$-spectra with BGBW distribution function for high-$\rm{S_{0}}$ (blue circle) and low-$\rm{S_{0}}$ (red triangle) in Pb-Pb collisions at $\rm{\sqrt{s_{NN}}}$ = 5.02 TeV. Linear fits for high-$\rm{S_{0}}$ and low-$\rm{S_{0}}$ are shown in dotted blue and red lines, respectively. The shaded area shows the uncertainties from the simultaneous BGBW fits.}
\label{betavstsphero}
\end{center}
\end{figure}

Figure~\ref{betavstsphero} shows the variation of kinetic freeze-out temperature ($\rm{T_{kin}}$) versus mean transverse radial flow velocity ($\rm{\langle\beta_{T}\rangle}$) for different spherocity and centrality classes obtained from simultaneous fit of identified charged particles' $p_{\rm{T}}$ spectra with BGBW function for Pb-Pb collisions at $\rm{\sqrt{s_{NN}}}$ = 5.02 TeV. The shaded area shows the uncertainties from the simultaneous BGBW fits. As we move from central to peripheral collisions, $\rm{T_{kin}}$ increases and $\rm{\langle\beta_{T}\rangle}$ decreases for all the spherocity classes. This behavior is naively expected due to the fact that central collisions are expected to have higher multiplicity in the final state, which would require more time to reach the freeze-out than the peripheral collisions. Conversely, with the increase in multiplicity one would expect higher transverse radial flow, which we observe in Fig.~\ref{betavstsphero}.  The spherocity dependent values of $\rm{\langle\beta_{T}\rangle}$ and $\rm{T_{kin}}$ for all centralities for Pb-Pb collisions at $\rm{\sqrt{s_{NN}}}$ = 5.02 TeV are enlisted in Table \ref{tab:freezeoutprop}.

\begin{table*} [!hpt]
                \centering
                \scalebox{1}{
                \begin{tabular}{|c |c |c | c| c | c| c|}
                \hline
                \multirow{2}{*}{Centrality(\%)} & \multicolumn{2}{c|}{{\textbf{High-$S_{0}$}}} & \multicolumn{2}{c|}{{\textbf{$S_{0}$ Integrated}}} & \multicolumn{2}{c|}{{\textbf{ Low-$S_{0}$}}} \\ \cline{2-7} 
                & $\rm{\langle\beta_{T}\rangle}$ & $\rm{T_{kin}}$ [GeV] & $\rm{\langle\beta_{T}\rangle}$ & $\rm{T_{kin}}$ [GeV] & $\rm{\langle\beta_{T}\rangle}$ & $\rm{T_{kin}}$ [GeV]  \\
             	\hline
                \hline
                0--10	&0.539 $\pm$ 0.007	 &0.124 $\pm$ 0.002	 &0.543 $\pm$ 0.006	 &0.123 $\pm$ 0.002	 &0.545 $\pm$ 0.007	 &0.124 $\pm$ 0.002 \\
                \hline
                10--20	&0.550 $\pm$ 0.007	 &0.123 $\pm$ 0.002	 &0.549 $\pm$ 0.006	 &0.123 $\pm$ 0.002	 &0.540 $\pm$ 0.008	 &0.129 $\pm$ 0.003 \\
                \hline
                20--30	&0.549 $\pm$ 0.006	 &0.122 $\pm$ 0.002	 &0.553 $\pm$ 0.006	 &0.122 $\pm$ 0.002	 &0.549 $\pm$ 0.007	 &0.128 $\pm$ 0.003 \\
                \hline
                30--40	&0.548 $\pm$ 0.007	 &0.120 $\pm$ 0.002	 &0.546 $\pm$ 0.006	 &0.123 $\pm$ 0.002	 &0.542 $\pm$ 0.007	 &0.128 $\pm$ 0.003 \\
                \hline
                40--50	&0.538 $\pm$ 0.006	 &0.121 $\pm$ 0.002	 &0.535 $\pm$ 0.006	 &0.125 $\pm$ 0.002	 &0.532 $\pm$ 0.007	 &0.130 $\pm$ 0.002 \\
                \hline
                50--60	&0.519 $\pm$ 0.007	 &0.125 $\pm$ 0.002	 &0.515 $\pm$ 0.007	 &0.129 $\pm$ 0.002	 &0.514 $\pm$ 0.008	 &0.134 $\pm$ 0.003 \\
                \hline
                60--70	&0.492 $\pm$ 0.008	 &0.129 $\pm$ 0.003	 &0.484 $\pm$ 0.008	 &0.134 $\pm$ 0.003	 &0.483 $\pm$ 0.009	 &0.139 $\pm$ 0.003 \\
                \hline
                70--100	&0.437 $\pm$ 0.010	 &0.136 $\pm$ 0.003	 &0.407 $\pm$ 0.015	 &0.145 $\pm$ 0.007	 &0.307 $\pm$ 0.021	 &0.163 $\pm$ 0.008 \\
                \hline
                \end{tabular} 
                } 
                \caption{ Kinetic freeze-out temperature ($\rm{T_{kin}}$) and mean transverse radial flow velocity ($\rm{\langle\beta_{T}\rangle}$) obtained from simultaneous fit of identified charged particles’ $p_{\rm{T}}$-spectra with Boltzmann-Gibbs blastwave function. 
                \label{tab:freezeoutprop}}
\end{table*}

Contrary to the observables studied so far in the previous sections, we see a clear dependence of the kinetic freeze-out parameters on $\rm{S_{0}}$. An isotropic (high-$\rm{S_{0}}$) event is expected to be dominated by large number of soft particles, which would require more time to reach the freeze-out. Thus, the isotropic events are found to have lower $\rm{T_{kin}}$ compared to  low-$\rm{S_{0}}$ events. To understand the dependence of the kinetic freeze-out parameters, we have fitted the parameters with a first order polynomial function separately for each spherocity classes. As evident from Fig.~\ref{betavstsphero}, the role of spherocity plays a bigger role in peripheral events and it diminishes when one goes towards central collisions. This could be an indication that spherocity is crucial while studying the final state effects.

\section{Summary and Conclusion}
\label{section4}
We have implemented transverse spherocity in Pb-Pb collisions at  $\sqrt{s_{\rm NN}}$ = 5.02 TeV using AMPT model and study the dependence of transverse spherocity on different global observables in heavy-ion collisions at the Large Hadron Collider energies. In summary, the spherocity distributions in Pb-Pb collisions are found to be shifted more towards the isotropic limit when compared to the pp collisions, where the distributions are shifted towards the jetty limit. This behavior is understood based on the fact that the system size in Pb-Pb collisions are significantly higher when compared with pp collisions and because of the medium effects through the process of
isotropisation, the jettiness of the events gets suppressed to a larger extent. The Bjorken energy density and speed of sound are found to be independent of the spherocity selection in heavy-ion collisions. However, we found that kinetic freeze-out parameters depend on spherocity. The role of spherocity plays a bigger role in peripheral events and it diminishes when one goes towards central collisions. The sensitivity
of event topology however, depends on the observable under study because of some of the counter-balancing effects in view of medium effects in heavy-ion
collisions or high-multiplicity environments.


\bibliography{sample}

\section*{Acknowledgements}

RS acknowledges the financial supports under the CERN Scientific Associateship and the financial grants under DAE-BRNS Project No. 58/14/29/2019-BRNS of Government of India. DB acknowledges the financial supports from CSIR, Government of India. SP acknowledges the financial supports from UGC, Government of India. ST acknowledges the supports under the INFN postdoctoral fellowship. The authors would like to acknowledge the usage of resources  of the LHC grid computing facility at VECC, Kolkata and usage of resources of the LHC grid Tier-3 computing facility at IIT Indore. The authors acknowledge Dr. Arvind Khuntia and Mr. Dushmanta Sahu for careful reading of the manuscript.

\section*{Author contributions statement}
All authors contributed equally to the conceptualization of the problem, event generation, data analysis, interpretation of the results, 
and manuscript preparation and reviewing.


\end{document}